\documentclass[
reprint,
superscriptaddress,
 amsmath,amssymb,
 aps,
]{revtex4-2}

\usepackage[pdftex]{graphicx}
\usepackage{dcolumn}
\usepackage{bm}
\usepackage{braket}
\usepackage{times}
\usepackage[mathlines]{lineno}

\allowdisplaybreaks[1]

\begin{document}

\preprint{APS/123-QED}

\title{Half-mirror for electrons on quantum Hall copropagating edge channels}

\author{Takase Shimizu}
\affiliation{
 Institute for Solid State Physics, The University of Tokyo, 5-1-5 Kashiwanoha, Kashiwa, Chiba 277-8581, Japan}
\author{Jun-ichiro Ohe}
\affiliation{
 Department of Physics, Toho University, 2-2-1 Miyama, Funabashi, Chiba, 274-8510, Japan}
\author{Akira Endo}
\affiliation{
 Institute for Solid State Physics, The University of Tokyo, 5-1-5 Kashiwanoha, Kashiwa, Chiba 277-8581, Japan}
\author{Taketomo Nakamura}
\affiliation{
 Institute for Solid State Physics, The University of Tokyo, 5-1-5 Kashiwanoha, Kashiwa, Chiba 277-8581, Japan}
\author{Shingo Katsumoto}
\affiliation{
 Institute for Solid State Physics, The University of Tokyo, 5-1-5 Kashiwanoha, Kashiwa, Chiba 277-8581, Japan}

\date{\today}

\begin{abstract}
A half-mirror that divides a spin-polarized electron into two parallel copropagating spin-resolved quantum Hall edge channels one half each is presented in this study.
The partition process was coherent, as confirmed by observing the Aharonov--Bohm oscillation at a high visibility of up to 60\% in a Mach--Zehnder interferometer, which comprised two such half-mirrors.
The device characteristics were highly stable, making the device promising in the application of quantum information processing.
The beam-splitting process is theoretically modelled, and the numerical simulation successfully reproduces the experimental observation.
The partition of the electron accompanied by the spin rotation is explained by the angular momentum transfer from the orbital to the spin via spin--orbit interactions.
\end{abstract}

\maketitle


\section{\label{sec:level1}Introduction}
Owing to their chiral nature,  quantum Hall edge channels (QHECs) present a significantly  high quantum coherence, as evidenced by the operations of the Mach--Zehnder interferometers (MZIs) \cite{Ji2003}, and Fabry--Perot interferometers \cite{Nakamura2019, Nakamura2020}, which both revealed fractional statistics of quasiparticles.
The high quantum coherence in QHECs has expanded the field of ``electron quantum optics," in which the one-dimensional edge states work as quantum beams.
These quantum beams have gained attention not only from physicists as fermionic beams with many-body interactions \cite{PhysRevLett.108.196803}, but also from researchers relative to quantum information processing \cite{Yamamoto2012}. Considering the latter, the electron quantum optical circuits have the following twofold roles: being the carrier of quantum information, ``wiring" clusters of qubits, and processing the quantum information within themselves \cite{PhysRevLett.84.5912}.
Here, we focus on the latter, where quantum gate operations are required on the beams of the circuits.
Several experimental and theoretical efforts have been made to construct building blocks of quantum circuits with QHECs \cite{B_uerle_2018, Hermann2022}, such as single-electron sources \cite{Feve1169, Giblin2012, kataoka2016, Ubbelohde2015}, detectors \cite{PhysRevLett.124.127701}, and controlled phase shifters \cite{PhysRevB.102.035417}.
These electron quantum beams have an orbital (charge) degree of freedom and spin degree of freedom.
When the spin--orbit interaction (SOI) is weak, two paths with quantum tunneling should be prepared for the orbital operation \cite{Yamamoto2012}, whereas a single path is sufficient for the spin, which is the internal degree of freedom \cite{Hermelin2011,Sanada2013}.
In both ways, each electron traveling on the track of the beams can be viewed as a flying qubit (FQ), which carries quantum information of one qubit.

\begin{figure*}
  \includegraphics[width=1\linewidth]{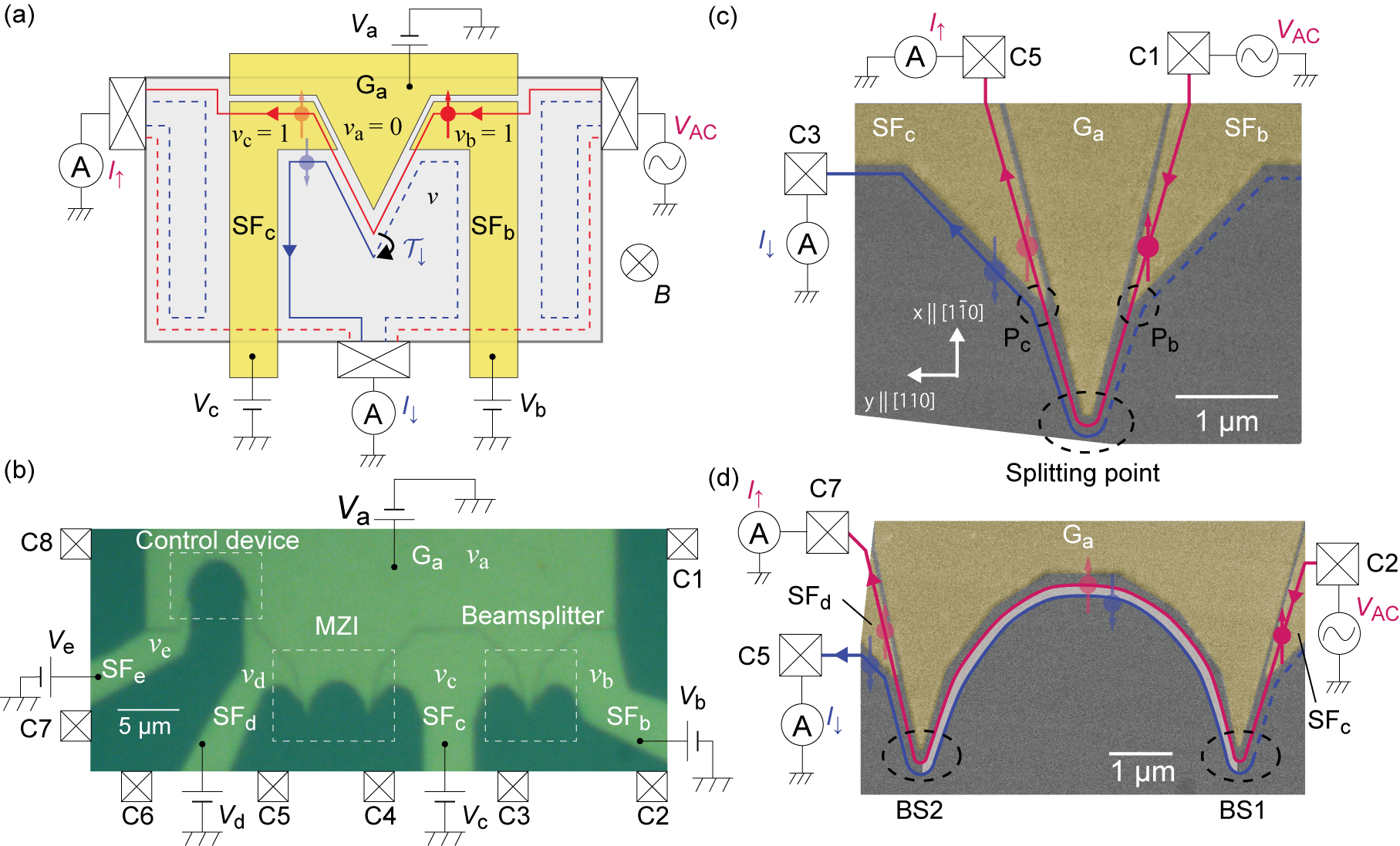}
  \caption{(a) Schematic of the BS  device. Channels 1 and 2 are indicated by the red and blue lines, respectively. (b) Optical micrograph of the sample with the gate and ohmic contact configuration. Brighter regions indicate the five metallic gates annotated as gate ${\rm G_a}$ and spin-filter gates ${\rm SF_{b, c, d, e}}$. Filling factors under the gates are annotated as $\nu_{\rm a, b, c, d, e}$.
(c) Magnified view of the BS obtained by a scanning electron microscope (Gates are yellow).
The parameters were set to $\nu_{\rm a}=\nu_{\rm d}=0$ and $\nu_{\rm b}=\nu_{\rm c}=1$.
For the measurement of ${\cal T}_\downarrow$, $V_{\rm AC}$ was applied to contact C1, and the currents at contact C5 and C3 were measured as $I_\uparrow$ and $I_\downarrow$, respectively.
(d) Magnified view of the MZI. The settings for the interferometry are $\nu_{\rm c}=\nu_{\rm d}=1$, $\nu_{\rm b}=\nu_{\rm e}=0$; $V_{\rm a}$ was applied in the range of $\nu_{\rm a}=0$.
For the measurement of the interference, $V_{\rm AC}$ was applied to contact C2, and the current at contacts C7 and C5 were measured as $I_\uparrow$ and $I_\downarrow$, respectively.}
\label{fig:SamplePicture}
\end{figure*}

Half-mirrors or beam-splitters (BSs) with a 1:1 ratio \cite{Yamamoto2005} are essential for quantum optical operations.
As electrical BSs, quantum point contacts (QPCs), which partially transmits electrons and reflects the residual part to a {\it counterpropagating} QHEC, have been utilized in most experiments performed thus far \cite{PhysRevLett.60.848}.
This property stems from the chiral nature of QHECs that sustains the coherence and is unavoidable for the QPC-BS scheme \cite{PhysRevB.38.9375}. 
The transition probability is controllable with a gate voltage of up to 50\%, at which the QPC functions as a Hadamard gate \cite{DiVincenzo2000,Ionicioiu2001}.

Although the QPC-BS scheme has several advantages including controllability, it presents certain drawbacks. For instance,
the area occupied by the circuit unit is inevitably large owing to the counterpropagating nature of the split beams.
Generally, the scalability of the solid-state circuit is a key advantage over optical quantum circuit.
However, the large area consumption in the QPC-BS scheme largely reduces the scalability.
Furthermore, the area consumption also leads to instability in quantum gate operations because the Aharonov--Bohm (AB) phase, which is a critical parameter in the qubit operation, fluctuates with the fluctuation of the external field and the gate voltages determining the area \cite{Ji2003}.
Another problem related to the instability is a type of cross-talk,
that is, the QPC partition ratio tends to be affected by the surrounding gate operations of the qubits.
The QPC should function in a transition region between conductance plateaus, where the ratio is sensitive to the small electric fields, as indicated by its application to single-charge sensors\cite{doi:10.1063/1.2794995}.
We also highlighted that the freedom in designing quantum circuits with QPC-BSs is limited due to topological constraints, such as the complex  series concatenation of MZIs.

The BS device described in this study functions between {\it copropagating} spin-split QHECs.
The spin-splitting of QHECs is caused by the Zeeman effect enhanced by the exchange of coupling between the electron spins, which lines up the spins in parallel inside the channels.
The series connection of multiple quantum gates is available \cite{PhysRevB.77.155320}, thus providing extensive freedom in the circuit design.
The distances between spin-split QHECs are generally short,  and the circuit areas can be reduced by decreasing the  distances between consecutive BSs, thus enhancing the stability of circuits.
Here, a beam-splitting action, such as partial interchannel tunneling, is a challenge that should be associated with spin flipping.
Previous studies regarding this type of tunneling transition used current imbalance \cite{Deviatov_2012, PhysRevB.77.161302, PhysRevB.84.235313}, and periodic magnetic gates with up to 28\% of transition probability \cite{PhysRevLett.107.236804, PhysRevB.92.195303}.
Nakajima {\it et al.} \cite{Nakajima2012, Nakajima2013} reported that spin-flips occurred at corners of the QHECs because the effective magnetic field of a spin--orbit interaction \cite{Frolov2009} changes nonadiabatically.
We also recently reported spin-flips at corners: the transition probability can be controlled by modulating the curvature of the QHECs by controlling the electrostatic potential via gate voltages \cite{PhysRevB.102.235302}.
However, the estimated transition probability of the BS was only up to 2\%, which is significantly below 50\% (half-mirror condition). Thus, the aforementioned absolute criterion of quantum information processing was not fulfilled.
Considering the geometrical effects \cite{PhysRevB.45.9059, PhysRevB.97.205419} and spin-rotation mechanism \cite{PhysRevB.102.235302}, a higher transition probability should be obtained by sharpening the corner, that is, enhancing  the curvature.

This study presents a half-mirror for copropagating spin-split QHECs with an acute-angle gate, which gains a high probability in nonadiabatic transition via SOI.
The transition probability can be controlled from 0\% to above 50\% with gate voltages by modulating the spatial distance between the channels at the bending point.
A high quantum coherence over the transition is confirmed by observing the AB interference in the MZI composed of two series BSs.
We observed a visibility of approximately 60\%, which is probably the highest among the previously reported MZIs using copropagating QHECs.
\section{Experimental methods and results}
\textit{Method}.--Figure \ref{fig:SamplePicture} (a) presents a schematic of the BS device fabricated on a two-dimensional electron system (2DES) in the spin-split integer quantum Hall regime.
The device consists of three Schottkey gates: gate ${\rm G_{\rm a}}$ is for beam splitting; and gates ${\rm SF_{\rm b}}$ and ${\rm SF_{\rm c}}$ are for spin filtering.
The external circuits and QHECs are illustrated in the figure.
Only QHECs with the spin-resolved Landau indices of $j=1$ and $2$ are depicted in the figure, although the filling factor $\nu=4$ in the nongated region was used in this experiment. The channels with $j=3$ and $4$ lying in the interior of the 2DES are omitted for simplicity.
We focus on the two QHECs, denoted as channel 1 (red) and channel 2 (blue), in which spins were locked at $\uparrow$ and $\downarrow$, respectively.
Beneath the spin-filter gates denoted as ${\rm SF_{\rm b}}$ and ${\rm SF_{\rm c}}$, the filling factors $\nu_b$ and $\nu_c$ were tuned to 1 so that only channel 1 proceeds through them \cite{Hashisaka2017, HASHISAKA201832}.
Gate ${\rm G_{\rm a}}$ was tuned to deplete the electrons under the gate.
In this configuration, a wave packet injected into channel 1 from the right electrode at the bias voltage $V_{\rm AC}$ meets channel 2 at the left corner of gate ${\rm SF_{\rm b}}$, and propagates to the right corner of gate ${\rm SF_{\rm c}}$.
The partial transition from channel 1 to 2 occurs through a local SOI \cite{PhysRevB.102.235302} as a result of the orbits wrapping around the sharp corner of gate ${\rm G_{\rm a}}$, annotated as ``Splitting point." The transition ratio is detected as the ratio of the currents through the left and bottom contacts, $I_{\uparrow}$ and $I_{\downarrow}$, respectively.
Based on the Landauer-B\"uttiker formula \cite{PhysRevB.38.9375}, the transition probability at the BS to channel 2 is given by ${\cal T}_\downarrow=G_\downarrow/(e^2/h)$, where $G_\downarrow \equiv I_\downarrow/V_{\rm AC}$. $e^2/h$ is the quantized conductance with an electron charge $e$ and Planck constant $h$.

A two-dimensional electron system (2DES) with an electron density of $3.8\times 10^{11}~{\rm cm^{-2}}$ and a mobility of $90~{\rm m^2/Vs}$ in a ${\rm Al}_x{\rm Ga}_{1-x}{\rm As}/{\rm GaAs}~(x=0.265)$ single heterostructure was used as a base system for the sample.
The structure of the wafer (from the front surface) consists of a 
5~nm Si-doped GaAs cap layer, 40~nm Si-doped ($N_{\rm Si} = 2 \times 10^{18}$~$\rm{cm}^{-3}$) ${\rm Al}_x{\rm Ga}_{1-x}{\rm As}$ layer, 15~nm undoped ${\rm Al}_x{\rm Ga}_{1-x}{\rm As}$ spacer layer, and an 800~nm GaAs layer with a 2DES residing near the interface with the upper layer.
Figure~\ref{fig:SamplePicture}(b) presents an optical micrograph of the sample with the configuration of the Au/Ti gates and ohmic contacts.
Three devices, including the beam-splitter, interferometer, and control device, were fabricated on a single substrate. The magnified views (from a scanning electron micrograph) of the beam-splitter and the MZI are shown in Fig.~\ref{fig:SamplePicture} (c) and (d), respectively.
The crystal orientation depicted in Fig.~\ref{fig:SamplePicture} was chosen to enlarge the rotation angle of the sum of the effective fields of the Rashba and Dresselhaus SOI at the corner \cite{Kunihashi2016}.

\begin{figure}
\includegraphics[width=1\linewidth]{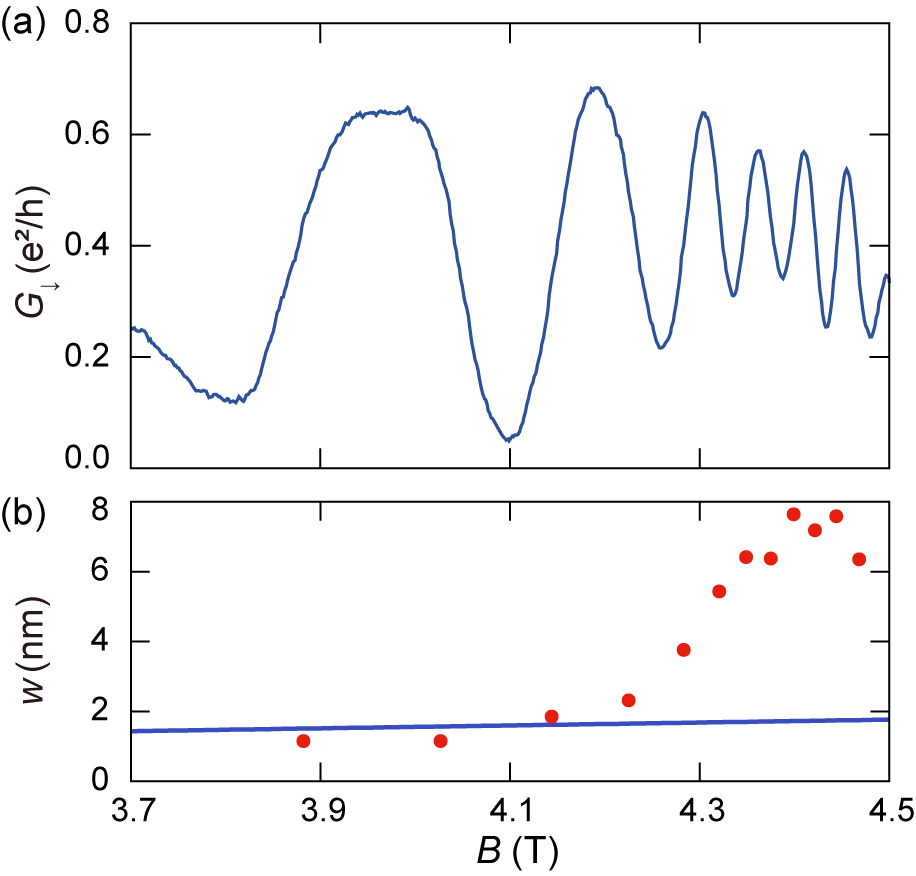}
\caption{(a) $G_\downarrow$-output of the MZI device on $\nu=4$ plateau as a function of $B$.
The gate voltages are set to achieve the channels, as illustrated in Fig.~\ref{fig:SamplePicture}(d).
$V_{\rm a}$ on G$_{\rm a}$ is fixed to $-$0.75~V.
$V_{\rm AC}=10~{\rm \mu V_{\rm rms}}$.
(b) Red dots indicate the width of the incompressible strip $w$ obtained from the oscillation data as $w=(2/3)(h/e)/L\Delta B$ at $V_{\rm a}=-0.75~{\rm V}$, where $\Delta B$ is estimated to be twice the distance between the adjacent oscillation peak and dip, and $L=8.28~{\rm \mu m}$.
The blue line indicates $w$ calculated from Eq.~(\ref{Larkin}) with the following parameters: $g^*=-0.6$ \cite{PhysRevB.102.235302}; $\epsilon=12.35$ \cite{doi:10.1063/1.88755}.}
\label{fig:MZ-075V}
\end{figure}

We cooled the sample down to 35~mK and applied a perpendicular magnetic field $B$ up to 8~T,
at which the 2DES was in the quantum Hall state with a filling factor of $\nu=2$.
A typical AC voltage of $V_{\rm AC}=25~{\rm \mu V_{\rm rms}}$ (except for the measurements in Fig.~\ref{fig:MZ-075V}) was applied at 170 Hz, and
the current was measured with a transimpedance amplifier using the standard lock-in technique.
We first measured the gate voltage dependence of the two-terminal conductance of the spin-filter gates ${\rm SF_{\rm b, c, d, e}}$ at several magnetic fields, as provided in the supplemental material (SM) \cite{Supplement2022}.
The gate voltages for the spin-filters were set around the centers of $e^2/h$ plateaus.

\begin{figure}
\includegraphics[width=\linewidth]{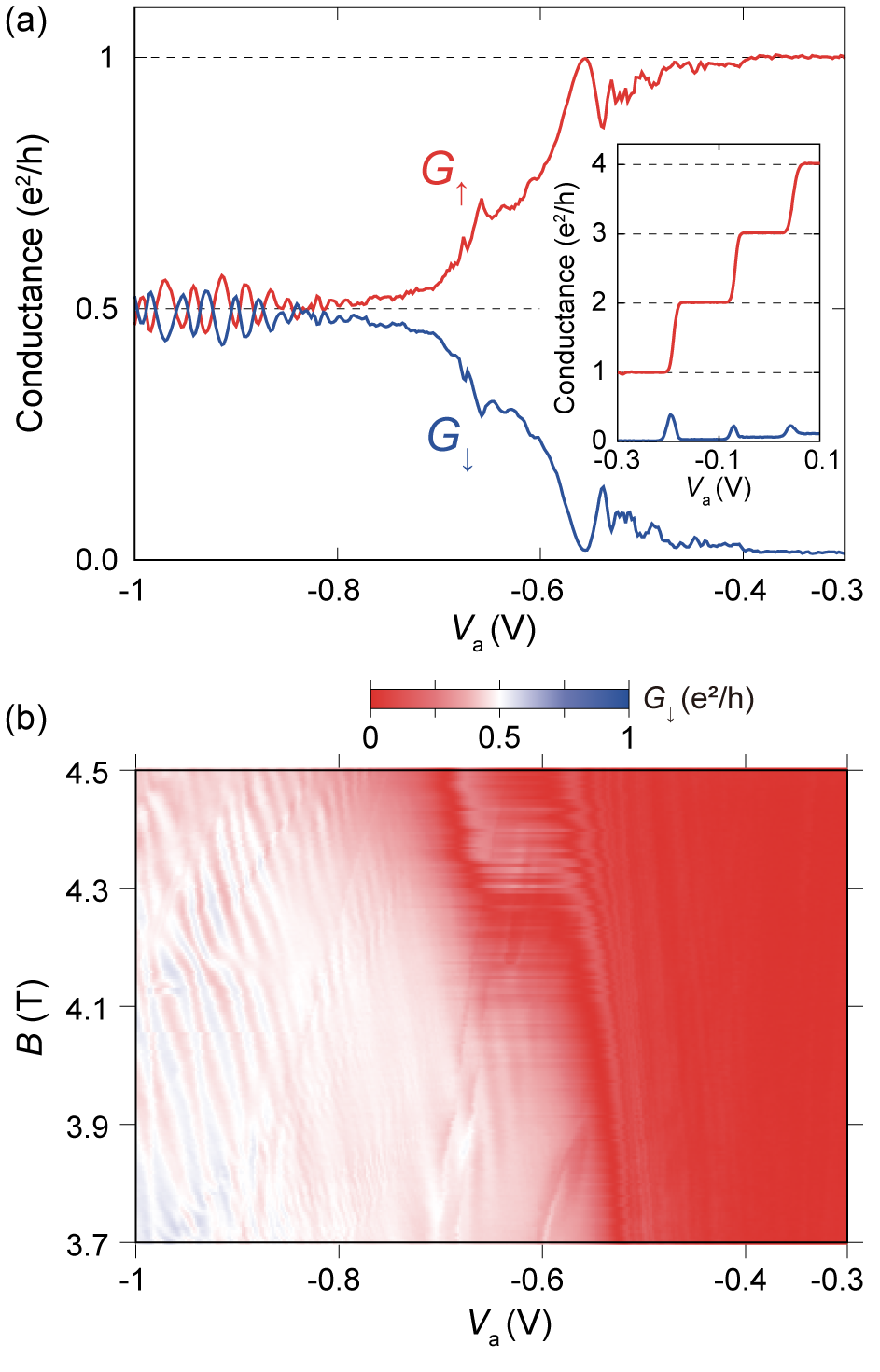}
\caption{Outputs of the beam-splitter.
The data were obtained in the configuration shown in Fig.~\ref{fig:SamplePicture}(c) with $V_{\rm b}=V_{\rm c}=-0.28~{\rm V}$ ($\nu_{\rm b}=\nu_{\rm c}=1$) and $V_{\rm d}=-0.5~{\rm V}$ ($\nu_{\rm d}=0$).
(a) $V_{\rm a}$-dependence of $G_{\uparrow,\downarrow}\equiv I_{\uparrow,\downarrow}/V_{\rm AC}$ at $B=4.1~{\rm T}$ ($\nu=3.8$).
The inset presents the data from $V_{\rm a}=0.1~{\rm V}$ to $-0.3~{\rm V}$.
(b) Color plot of $G_\downarrow$ as a function of $V_{\rm a}$ and $B$.}
\label{fig:BS-1D2D}
\end{figure}

\textit{Beam-splitter}. -- The experimental results of the BS device are presented herein.
As illustrated in Fig.~\ref{fig:SamplePicture}(a), we first set $\nu_{\rm b}=\nu_{\rm c}=1$ by applying $V_{\rm b}=V_{\rm c}=-0.28~{\rm V}$.
Then, we changed $V_{\rm a}$ from $0.1~{\rm V}$ to $-1~{\rm V}$.
Figure~\ref{fig:BS-1D2D}(a) demonstrates the $V_{\rm a}$-dependence of $G_{\uparrow,\downarrow}$ at $B=4.1~{\rm T}~(\nu=3.8)$.
The inset in Fig.~\ref{fig:BS-1D2D}(a) presents the data in the region from $V_{\rm a}=0.1~{\rm V}$ to $-0.3~{\rm V}$, where four plateaus corresponding to $\nu_{\rm a}=4, 3, 2, 1$ are observed.
This manifests that there were two extra QHECs inside the 2DES, other than channels 1 and 2, as shown in Fig.~\ref{fig:SamplePicture}(a).
However, the tunneling rate from channels 1 and 2 to the extra QHECs was significantly low (less than 2\%, as indicated in SM),
thus the two additional channels will be ignored in subsequent discussions.
When $V_{\rm a}$ was driven to be more negative than $\approx-0.4~{\rm V}$, $G_\uparrow$ started to decrease from $e^2/h$ and $G_\downarrow$ increased by the same amount. This indicates the division of the current into channels 1 and 2.
At $V_{\rm a}\approx-0.85~{\rm V}$, ${\cal T}_{\rm \downarrow}$ reached 50\%, and ${\cal T}_{\rm \downarrow}$ demonstrated an oscillatory behavior by further decreasing $V_{\rm a}$.
Because this critical value ($-$0.5~V) of $V_{\rm a}$ is near the voltage that corresponds to the establishment of the acute angle paths of channels 1 and 2 as shown below, the behavior in Fig.~\ref{fig:BS-1D2D}(a) indicates that the transmission was mostly around the acute angle, referred to as the ``Splitting point" (SP) in Fig.~\ref{fig:SamplePicture}(c).

\begin{figure}[tbh]
  \includegraphics[width=1\linewidth]{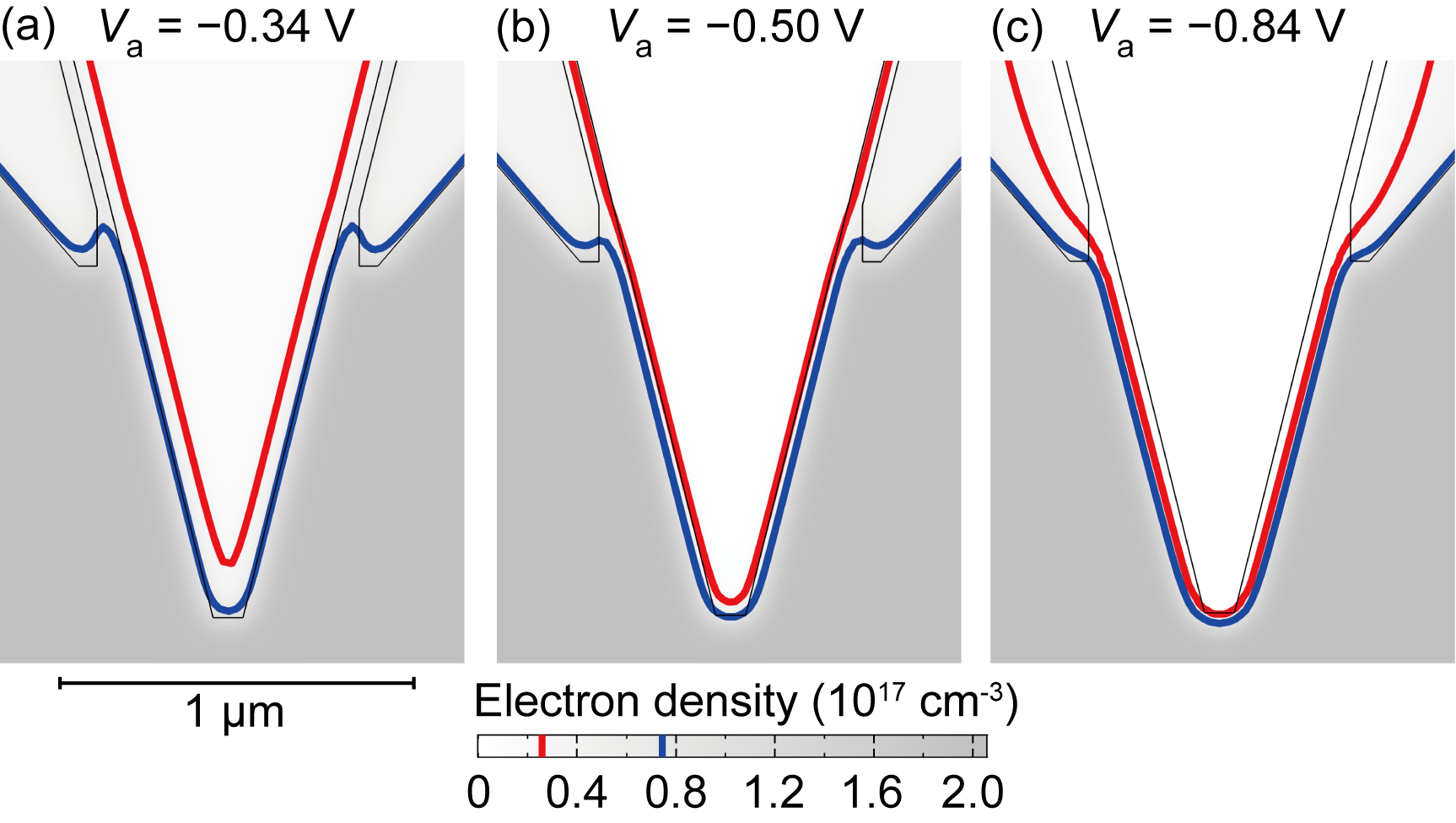}
  \caption{Grayscale plot of the electron density $N$ at a depth of $64~{\rm nm}$ from the surface, which was calculated with the finite-element method described in SM for various center gate voltages $V_{\rm a}$. 
The equipotential lines of $(1/2)(N_0/\nu)$ and $(3/2)(N_0/\nu)$, where $\nu=4$ and $N_0$ are the electron densities in the nongated region, are plotted in red and blue, indicating the position of channels 1 and 2, respectively.
 The voltage of the spin filters was set to $-0.28~{\rm V}$ in this simulation, which is the same as that in  the experiment. (a) Plot of $V_{\rm a}=-0.34~{\rm V}$. (b) Plot of $V_{\rm a}=-0.5~{\rm V}$. (c) Plot of $V_{\rm a}=-0.84~{\rm V}$.}
\label{Sup_fig_COMSOL}
\end{figure}
The aforementioned behavior of ${\cal T}_{\rm \downarrow}$ for $V_{\rm a}$ can be explained by considering the $V_{\rm a}$ dependence of the distance between channels 1 and 2 around the SP.
A numerical calculation using the finite-element method demonstrates that the equipotential lines around gate ${\rm G_a}$ intricately change against $V_{\rm a}$, as shown in Fig.~\ref{Sup_fig_COMSOL}.
Based on the two-terminal conductance of the spin-filters, most of the 2DES below gate ${\rm G_a}$ was determined to be depleted at $V_{\rm a}\approx-0.35~{\rm V}$, and channel 1 approached channel 2.
Nevertheless, $G_\downarrow$ remained zero in this region because $V_{\rm a}=-0.35~{\rm V}$ was insufficient to deplete electrons below the splitting point of ${\rm G_{\rm a}}$ owing to the narrow shape of the gate; thus, channels 1 and 2 remained spatially separated at the splitting point, as shown in Fig.~\ref{Sup_fig_COMSOL}(a).
$G_{\rm \downarrow}$ began to increase with a negative $V_{\rm a}$ at approximately $-$0.5~V, where the 2DES at the SP started to deplete, further leading the channel 1 to approach channel 2 and thus enhancing the transition rate.

Figure~\ref{fig:BS-1D2D}(b) presents a color plot of $G_{\uparrow}$ on the plane of $V_{\rm a}$ and $B$ in the $\nu=4$ plateau region.
The overall trend is that $G_{\uparrow}$, hence ${\cal T}_{\uparrow}$ decreased with an increase of $B$, reflecting an increase in the distance between channels 1 and 2 at the SP \cite{PhysRevB.102.235302}.
The weak oscillation of $G_{\uparrow,\downarrow}$ in the region $V_{\rm a}<-0.85$~V slowly shifted with $B$, indicating that this was owing to an AB interference with a significantly narrow area of the interference loop.
As shown in Fig.~\ref{fig:SamplePicture}(c), in addition to the SP, the beam-splitter has two bending points of channels 1 and 2 at the vertices of SF$_{\rm b}$ and SF$_{\rm c}$, noted as P$_{\rm b}$ and P$_{\rm c}$ respectively.
These two points may function as extra nodes of the electron beams and form weak interferometers combined with the SP.
As shown in the SM, the starting point of the oscillation in $V_{\rm a}$ shifted with $V_{\rm b}$, while it was insensitive to $V_{\rm c}$, indicating that P$_{\rm b}$ functioned as a weak node and P$_{\rm c}$ did not, although no clear explanation for the difference can currently be provided.

\textit{Mach--Zehnder interferometer}.--Observing the interference is a direct way to test the quantum coherence over the beam splitting process.
For the experiment of the MZI configuration displayed in Fig.~\ref{fig:SamplePicture}(d), we first set $\nu_{\rm c}=\nu_{\rm d}=1$ and $\nu_{\rm b}=\nu_{\rm e}=0$ for the two channles under consideration to run as illustrated.
The following was obtained from the Landauer-B\"uttiker formula: ${\cal T}_\downarrow=|r_1e^{i\phi_1}t_2+t_1e^{i\phi_2}r_2|^2=|t_1r_2|^2+|r_1t_2|^2-2|t_1t_2r_1r_2|\cos\phi$. 
Here, $t_i$ ($r_i$) is the transmission (reflection) amplitude of the $i$th BS fulfilling $|r_i|^2+|t_i|^2=1$, and \cite{Nakajima2012, PhysRevB.102.235302}
\begin{align}\label{Eq:ABphase}
\phi=\phi_1-\phi_2=2\pi \frac{w L B}{h/e}
\end{align}
is the AB phase,
where $w$ is the width of incompressible strip \cite{PhysRevB.46.4026,PhysRevB.52.R5535} between channels 1 and 2 averaged over the arc-like curve, and $L$ is the length of the arc-like curve. 
The visibility of the oscillation is defined as 
$\mathcal{V} \equiv {\cal T}_{\downarrow \rm max}-{\cal T}_{\downarrow \rm min}=(G_{\downarrow \rm max}-G_{\downarrow \rm min})/(e^2/h)$.

Figure~\ref{fig:MZ-075V}(a) presents the $B$-dependence of $G_{\rm \downarrow}$ at $V_{\rm a}=-0.75~{\rm V}$, where the single BS in Fig.~\ref{fig:SamplePicture}(c) had a 50\% transition rate.
Because the two vertices in Fig.~\ref{fig:SamplePicture}(d) have the same angle as that in Fig.~\ref{fig:SamplePicture}(c), we can expect that in ${\cal T}_{1,2}\equiv|t_{1,2}|^2 \approx 0.5$.
$G_\downarrow$ oscillates as a function of $B$, with a significantly high visibility of $\mathcal{V} \approx 60\%$ in the region from $B=$3.9~T to 4.1~T.
$\mathcal{V}$ decreased above 4.1~T, which is probably due to the shifts of ${\cal T}_{1,2}$ from the optimal condition.
 
The oscillation period significantly decreased above 4.2~T, which can be explained in the same manner as in \cite{PhysRevB.102.235302}.
In a classical electrostatic model of the QHECs \cite{PhysRevB.46.4026,PhysRevB.52.R5535}, $w$ is given by
 \begin{align}\label{Larkin}
  w
  \approx
  \sqrt{\frac{8|g^*\mu_{\rm B} B|\epsilon\epsilon_0}{\pi e^2(dn/dr)|_{r=r'}}},
 \end{align}
where $r'$ is the center position of the incompressible liquid strip between channels 1 and 2 from the edge of gate ${\rm G_a}$; $\epsilon \epsilon_0$ is the dielectric permittivity of the matrix semiconductor; $n(r)$ is the electron sheet density profile, and $g^*\mu_{\rm B} B$ is the exchange-aided Zeeman splitting, where $g^*$ is the effective Land\'e $g$-factor; and $\mu_{\rm B}$ is the Bohr magneton.
In Fig.~\ref{fig:MZ-075V}(b), the red dots indicate the values of $w$ that were experimentally estimated using equation $w=(2/3)(h/e)/L\Delta B$, where $\Delta B$ is the oscillation period in $B$ (see \cite{PhysRevB.102.235302} for the detail).
The blue line indicates the value of $w$ calculated using Eq.~\eqref{Larkin}, with which the estimated $w$ reasonably agrees below 4.2~T, though it starts to deviate above that.
The deviation most likely arises from the breakdown of our assumption $w \propto \sqrt{B}$ above 4.2~T.
The exchange interaction has been demonstrated to lead to a stronger $B$-dependence of $w$ \cite{PhysRevB.73.075331, PhysRevB.73.155314, PhysRevB.78.035340}.
Furthermore, $B\approx4.2~{\rm T}$ is the field where the edge state transforms from a spin-textured state to a spin-polarized state \cite{PhysRevB.82.201302, PhysRevLett.77.2061, PhysRevB.56.10383, PhysRevB.66.165308}, and the exchange enhancement should be enlarged \cite{doi:10.1143/JPSJ.37.1044}.

\begin{figure}[t]
  \includegraphics[width=1\linewidth]{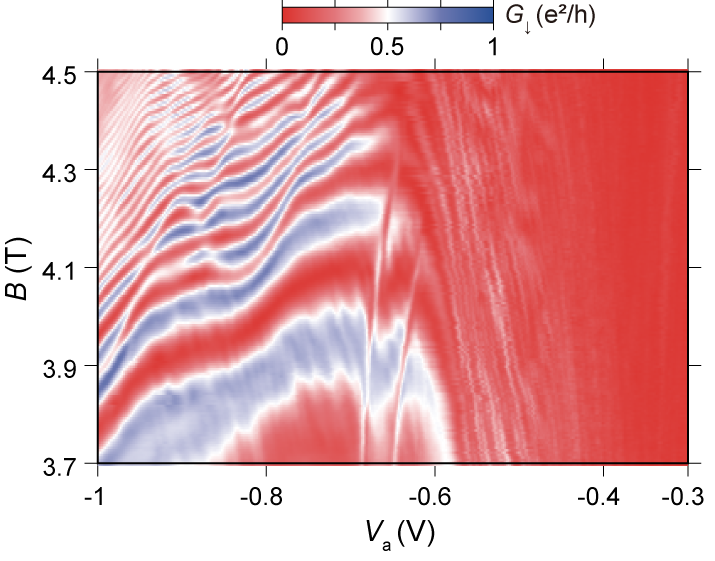}
  \caption{Color plot of measured $G_\downarrow$ as a function of $V_{\rm a}$ and $B$ obtained for the MZI.}
\label{fig:MZ2D}
\end{figure}
Figure~\ref{fig:MZ2D} presents a color plot of $G_\downarrow$ as a function of $V_{\rm a}$ and $B$.
The oscillation pattern appears as curved stripes below $V_{\rm a}\approx-0.6~{\rm V}$.
This arc-like pattern was also observed in a previous study \cite{PhysRevB.102.235302} and was successfully explained with the electrostatic treatment of QHECs \cite{PhysRevB.46.4026,PhysRevB.52.R5535}.
The apparent AB oscillation proves that the transition at the BS is highly coherent, eliminating the possibilities of other transition processes, that is, charge equilibration via impurity scattering \cite{PhysRevB.45.3932, PhysRevB.45.13777, PhysRevB.53.15777}.
Note, the visibility of 0.6 only indicates the lower boundary of coherence.
For the configuration of MZI in this study, the fine and independent tuning of the partition rates of the BSs was not possible, and it may have prevented the BS conditions from being tuned to the optimal point.
Although the BS is highly coherent, the results in the control device (SM) indicate that a certain amount of dephasing was caused by the inter-channel transition on the gate-defined curve.
The curvature of the beams during application should be as small as possible at the distance between the critical channels.

Note, the characteristics of the MZI were significantly stable and reproducible despite several weeks of measurements. 
The significantly small area enclosed by the two paths ($0.02{\rm \mu m^2}$ for $w=2~{\rm nm}$) prevents the smearing of the interference signal caused by the magnetic field fluctuation.
This high stability significantly contradicts to the MZIs with the BSs of QPC, which generally requires measurements in short periods \cite{Ji2003}. 
This is an advantage of the proposed scheme for the application of quantum information processing.

\section{Discussion and numerical simulation}
The aforementioned results are summarized as follows: 1) the transition rate of the BSs reached 50\%, and 2) the transition was highly coherent, as evidenced by the amplitude of the interference in the MZI.
Result 1) presents uncertainty regarding the picture demonstrated in Ref.~\cite{Nakajima2013}, where the rotation is caused by the abrupt change in the direction of the effective magnetic field, which is the external field plus the SOI field.
Because the estimated SOI field $B_{\rm SOI}\approx 1~{\rm T}$ in the GaAs/AlGaAs heterostructure \cite{PhysRevB.45.3932} is smaller than the external magnetic field $B=4.1~{\rm T}$, the rotation angle of the effective field at the corner should be less than $\pi/2$. Thus, the spin-flip rate cannot reach 50\% in this case.

Instead, the fact that the partition process indicates tunneling between the copropagating edge states at the same Fermi energy should be emphasized.
Here, the effective magnetic field presents no energy difference between the initial and final states.
On the other hand, a tunneling between states with opposite spins requires a variation in the angular momentum.
In this case, the only option for the source of angular momentum is the curved orbit around the transition point.
The SOI is indispensable for the angular momentum transfer.
Furthermore, within the Born approximation \cite{PhysRevB.85.155317}, the transition matrix for the spin-flip tunneling consists of the SOI terms.
Thus, the smaller the turning radius of the BS, the larger the spin rotation angle owing to the larger local angular momentum.

\begin{figure*}[bth]
\centering
\includegraphics[width=1\linewidth,angle=0]{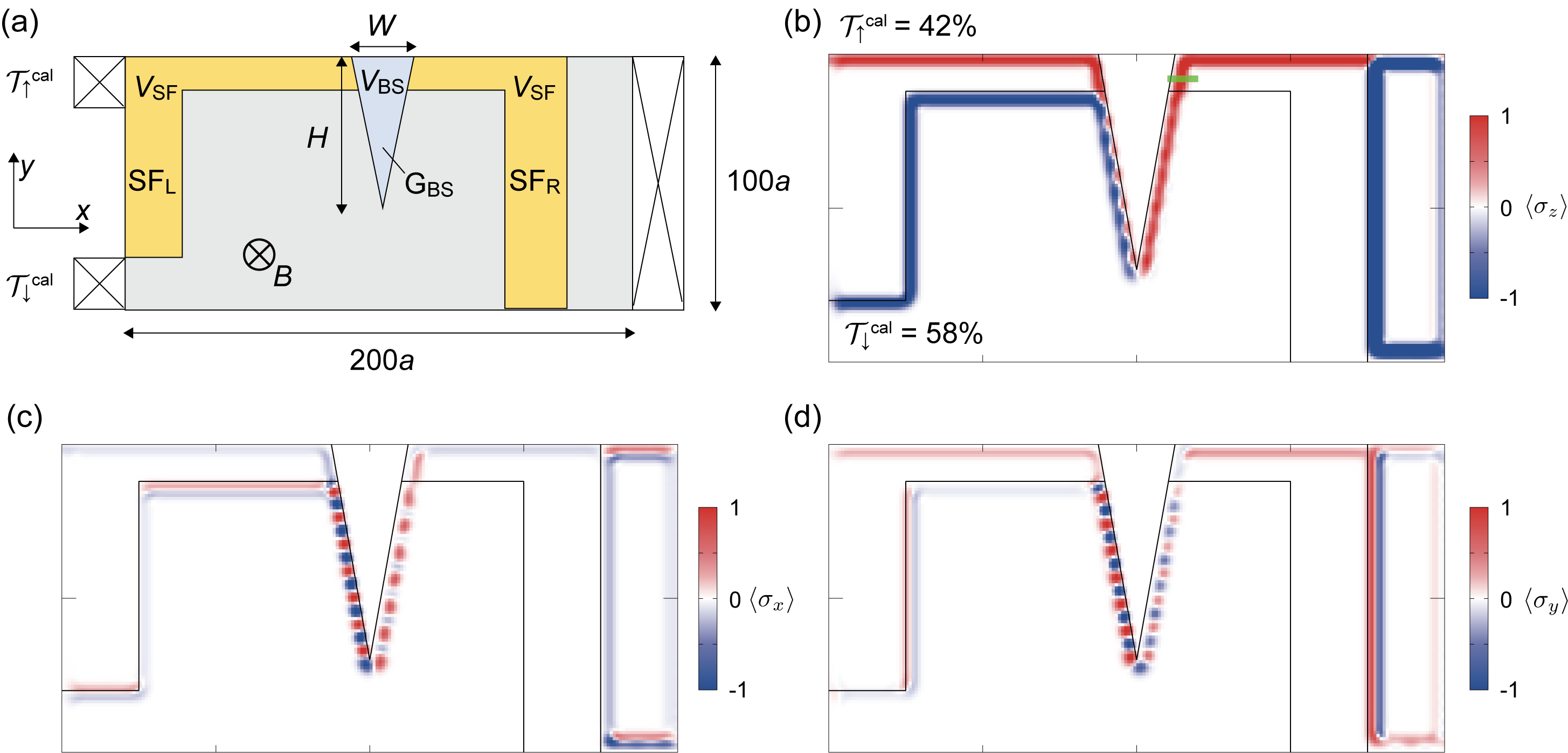}
\caption{(a)Schematic of the model used for the simulation of the BS.
Spin-filter gates ${\rm SF_R}$ and ${\rm SF_L}$ correspond to ${\rm SF_b}$ and ${\rm SF_c}$ in the experiment, respectively.
The potential is set to $V_{\rm SF}$ in the yellow regions.
The blue-gray triangular gate ${\rm G_{BS}}$ with potential $V_{\rm BS}$ corresponds to gate ${\rm G}_{\rm a}$ in the experiment.
Outgoing normal leads are attached to the left side of the sample.
(b) Calculated distribution of the $\sigma^z$ component $\left< \sigma^z \right>$ of the current carrying state, (c) $\sigma^x$ component $\left< \sigma^x \right>$, and (d) $\sigma^y$ component $\left< \sigma^y \right>$.}
\label{fig:Tight-binding-TypicalCondition}
\end{figure*}
\begin{figure*}[bth]
\centering
\includegraphics[width=1\linewidth,angle=0]{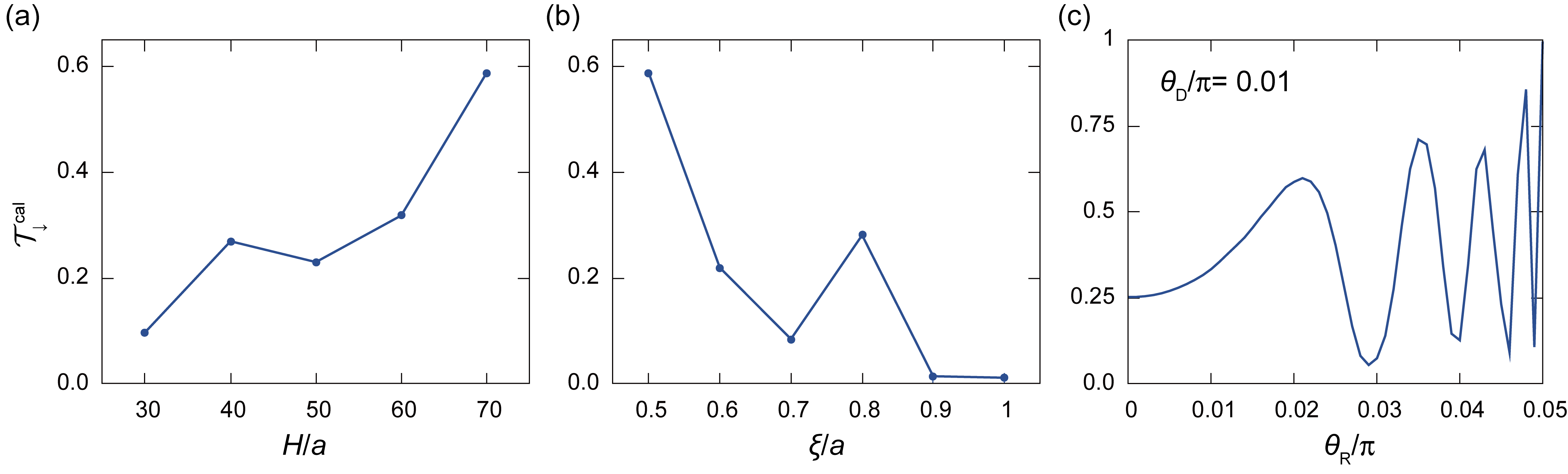}
\caption{Calculated transition probability ${\cal T}^{\rm cal}_\downarrow$ as a function of three different parameters. (a)${\cal T}^{\rm cal}_\downarrow$ as a function of the height $H$ of the triangular gate potential, with a fixed width of $W=25a$. The strength of SOI is set to $\theta_{\rm R}=0.02\pi$ and $\theta_{\rm D}=0.01\pi$. $\xi=0.5a$.
(b) ${\cal T}^{\rm cal}_\downarrow$ as a function of the slope of the gate potential $\xi$. Other parameters are as follows: $H=70a$, $W=25a$, $\theta_{\rm R}=0.02\pi$, $\theta_{\rm D}=0.01\pi$. (c) ${\cal T}^{\rm cal}_\downarrow$ as a function of the strength of the Rashba SOI $\theta_{\rm R}$. Other parameters are as follows: $H=70a$, $W=25a$, $\theta_{\rm D}=0.01\pi$, and $\xi=0.5a$.}
\label{fig:Tight-binding-graph}
\end{figure*}

\textit{Numerical simulation}.--To examine the aforementioned intuitive inference, the electron transport was numerically studied using the QHECs with the SOI.
The schematic of the calculated system is shown in Fig.~\ref{fig:Tight-binding-TypicalCondition} (a).
The Hamiltonian is as follows:
\begin{equation}
\mathcal{H}=\frac{{\bf \Pi}^2}{2m^*}+V({\bf r})+H_{\rm R}+H_{\rm D}+E_z\sigma^z,
\end{equation}
where ${\bf \Pi}=-i\hbar{\bf \nabla}+e{\bf A}$ is the kinetic momentum, $V({\bf r})$ is the potential forming the waveguide, $H_{\rm R}=(\alpha/\hbar)\left( \Pi_x \sigma^y - \Pi_y \sigma^x \right)$ and $H_{\rm D}=(\beta/\hbar)\left( \Pi_x \sigma^x - \Pi_y \sigma^y \right)$ are the Rashba and Dresselhaus type spin-orbit coupling, respectively, and $E_z$ is the Zeeman energy.
We adopted the tight-binding approximation and revised the Hamiltonian as follows:
\begin{equation}
\mathcal{H}=\sum_{i,\sigma} \left( V_i+E_z\sigma^z_{\sigma \sigma} \right) c_{i\sigma}^\dagger c_{i\sigma} - \sum_{\left< i,j \right>, \sigma, \sigma'}h_{i\sigma,j\sigma'}c_{i\sigma}^\dagger c_{j\sigma'},
\end{equation}
where $i,j$ are the site indices and $\left< i,j \right>$ indicates the nearest neighbors.
The hopping energy $h_{i,j}$ was defined as follows:
\begin{eqnarray}
h_{i,i+\hat{x}}&=&-V_{0}e^{-2\pi i \frac{\varphi_i}{\varphi_0}}
\begin{pmatrix}
1 & -\theta_{\rm R}+i\theta_{\rm D} \\
-\theta_{\rm R}+i\theta_{\rm D} & 1 \\
\end{pmatrix},
\\
h_{i,i+\hat{y}}&=&-V_{0}
\begin{pmatrix}
1 & i\theta_{\rm R}-\theta_{\rm D} \\
i\theta_{\rm R}+\theta_{\rm D} & 1 \\
\end{pmatrix},
\end{eqnarray}
where $V_{0}=\hbar^2/2m^{*}a^2$, $a$ is the lattice constant, $\varphi_0$ is the magnetic flux quantum, and $m^{*}=0.067m_{\rm e}$ is the effective mass.
$i+\hat{x}$ ($i+\hat{y}$) denotes the nearest neighbor of $i$ along the $x$ ($y$) axis in the positive direction.
The coupling constants of the SOI, $\alpha, \beta$ are related to $\theta_{\rm R}, \theta_{\rm D}$ by $\alpha=2\theta_{\rm R}V_0 a$ and $\beta=2\theta_{\rm D}V_0 a$ for small $\alpha, \beta$ \cite{PhysRevB.72.041308}.
In the Landau gauge, the magnetic field $B$ is given by $\varphi_i=(y_i-1)\varphi$, where $\varphi=Ba^2$ is the magnetic flux per lattice cell.

The recursive Green's function method was employed to calculate the electron transport and distribution of the current-carrying state \cite{PhysRevLett.47.882,PhysRevB.44.8017}.
The magnetic flux and Fermi energy of the conduction electron were set to $\varphi=0.05\varphi_0$ and $E=-3.5 V_0$, respectively, to form two QHECs in one of the edges of the system.
By choosing the parameters $E_{\rm z}=0.15 V_0$ and $V_{\rm SF}=0.2 V_0$, the gate potentials ${\rm SF_{R}}$ and ${\rm SF_{L}}$ enabled the system to transmit through only one channel.
Due to the Zeeman splitting, each electron channel was spin-polarized, and the system acts as a spin filter.
The system was $200a \times 100a$ in size.
An additional gate $G_{\rm BS}$ was introduced to facilitate the beam splitting, which was triangular and characterized by the width $W$ and the height $H$.
The form of the gate potential $V_i$ gradually decreased from the edge of the gates as
$V=V_{\rm BS,SF}\exp(-r/\xi)$,
where $r$ is the distance from the edge of the gates and $\xi$ determines the slope of the potential.

Figure \ref{fig:Tight-binding-TypicalCondition}(b), (c), and (d) present the spatial distribution of three different components ($\sigma^z$, $\sigma^x$, and $\sigma^y$) of the current carrying state for a typical parameter set of $H=70a$, $W=25a$, $\xi=0.5a$, $\theta_{\rm R}=0.02\pi$, and $\theta_{\rm D}=0.01\pi$. Let us assume $a=5~{\rm nm}$ to compare the calculated model with the experiments.
Then, the area of the system is $1~\mu{\rm m} \times 0.5~\mu{\rm m}$, and the other parameters are as follows: $V_0=23~{\rm meV}$, $B=4.1~{\rm T}$, $g^*=E_z/\mu_{\rm B}B=14$, $\alpha=1.4 \times 10^{-11}~{\rm eVm}$, and $\beta=7.1\times 10^{-12}~{\rm eVm}$. 
Note, these parameters were tuned to clarify the essential legitimacy of our hypothesis and are not necessarily the same as those in the experiment.
Namely, the simulation does not include several significant factors, such as the Coulomb interaction and, thus, cannot be directly compared.
Instead, the  qualitative and half-quantitative (relationship between the amplitudes of the effective magnetic fields and the spin rotation angle) properties should be emphasized.

Fig.~\ref{fig:Tight-binding-TypicalCondition}(b) demonstrates that the gate ${\rm SF_R}$ functions as a spin filter, and only the up-spin current goes through the gate.
The clear spin-flip inter-channel transition is observed at the bottom corner of the gate ${\rm G_{BS}}$, validating our interpretation of the experimental results.
The calculated transition probability is ${\cal T}^{\rm cal}_{\downarrow}=58\%$ for this parameter set.
Note, the ratio of the SOI field to the external field $B_{\rm SOI}/B$ in this simulation is estimated to be 0.26 by assuming $B_{\rm SOI}/B\approx \left<\sqrt{\sigma_x^2+\sigma_y^2}/\sigma_z \right>$, where the expectation value is obtained along the green line denoted in Fig. \ref{fig:Tight-binding-TypicalCondition}(b).
Thus, despite the condition where $B_{\rm SOI}$ is smaller than $B$, the transition probability reaches 50\%, supporting our intuitive discussion.
After the transition, the spin-up and spin-down wavefunctions co-propagate along the left side of the gate ${\rm G_{BS}}$, demonstrating a periodic change in the $\left< \sigma^z \right>$, which is also observed in the $\left< \sigma^x \right>$ and $\left< \sigma^y \right>$ components shown in Figs.~\ref{fig:Tight-binding-TypicalCondition}(c) and (d), representing the spin precession.
The small spin-flip inter-channel transition occurs when the current leaves and re-enters the spin-filter gate, leading to an AB oscillation, which explains the small oscillation in ${\cal T}_\downarrow$ observed in the experiment, as shown in Figs.~\ref{fig:BS-1D2D}(a) and (b) below $V_{\rm a}\approx -0.9~{\rm V}$.

Figure~\ref{fig:Tight-binding-graph}(a) demonstrates ${\cal T}^{\rm cal}_\downarrow$ as a function of the height of the triangular gate potential $H$, with a fixed width of $W=25 a$.
${\cal T}^{\rm cal}_\downarrow$ is apparently enhanced by increasing $H$.
Therefore, the sharp angle of the vertex of the triangular gate potential can be concluded to enhance the spin-flip inter-channel transition.
We obtained a half mirror (${\cal T}_\downarrow=50\%$) for $H\cong65a$.

The experimental results in Fig.~\ref{fig:BS-1D2D} demonstrate that ${\cal T}^{\rm cal}_\downarrow$ increases by applying a larger negative gate voltage.
This situation can be modeled by changing the slope of the gate potential $\xi$.
Fig.~\ref{fig:Tight-binding-graph} (b) demonstrates ${\cal T}^{\rm cal}_\downarrow$ as a function of $\xi$.
${\cal T}^{\rm cal}_\downarrow$ increases as $\xi$ decreases, with a slight oscillatory variation.
The reason for the increasing ${\cal T}^{\rm cal}_\downarrow$ is as follows.
First, the larger the potential gradient (smaller $\xi$) at the position of the edge channel, the smaller the distance between the gate edge to the edge channels, leading to the larger curvature of the corner. This enhances the nonadiabaticity of the transition.
Second, as the potential gradient becomes larger, the width of the potential barrier between the spin-split edge channels becomes narrower, leading to an enhanced tunneling probability.
The oscillatory variation is most likely caused by the phase modulation of the AB interference when the potential gradient at the edge channels is varied.

Figure~\ref{fig:Tight-binding-graph}(c) demonstrates ${\cal T}^{\rm cal}_\downarrow$ as a function of $\theta_{\rm R}$ with $\theta_{\rm D}=0.01\pi$. 
${\cal T}^{\rm cal}_\downarrow$ increases from $\theta_{\rm R}=0$ to $0.02$, and oscillates for a larger $\theta_{\rm R}$, which is sufficiently explained by the Rabi oscillation caused by the rotating effective field $B_{\rm SOI}$ at the corner of the BS. 
Note, when the Hamiltonian does not contain the spin-orbit term ($\theta_{\rm R}=\theta_{\rm D}=0$), which is not shown in the figure, ${\cal T}_\downarrow$ becomes zero, indicating that the SOI is indispensable for the inter-channel transition.

To summarize the simulation, the numerical simulation based on the recursive Green's function method successfully reproduces the experimental results, not only qualitatively but half-quantitatively, particularly in the zenith angle rotation of spin.
The result strongly supports our inference regarding the spin-rotation mechanism of the SOI-aided inter-channel quantum tunneling.

\section{Concluding remark}
We investigated the BS on copropagating QHECs using a metal gate with an acute angle corner.
The interchannel transitions accompanied by spin flips are caused by SOIs and the orbital angular momentum created at the sharp corner of the QHECs.
The transition probability of the BS can be controlled from approximately 0\% to over 50\% by modulating the distance of the copropagating QHECs via the gate voltage.
We composed an MZI with the BSs, demonstrating a high visibility in the interference pattern of up to 60\%, indicating a high quantum coherence in the transition over the BS.
The device characteristics were very stable, reflecting its compactness.
To the best of our knowledge, this is the first achievement of the half-mirror condition (50\% partition) on copropagating QHECs.
The results present a convenient, stable, and scalable way of processing quantum information in the flying qubit scheme.

We thank Dr. Hashisaka Masayuki for the fruitful discussion.
This work was partly supported by JSPS KAKENHI Grant Numbers JP19H00652, 19K05253, 21H05016, and 21H01799.

\bibliography{beamsplitter}
\end{document}